\documentclass[aps,prl,reprint,groupedaddress,10pt,twocolumn,showpacs]{revtex4}
\usepackage{graphicx}

\def\8{\infty}
\def\oh{\frac{1}{2}}

\def\d{\partial}

\def\undertext#1{\vtop{\hbox{#1}\kern 1pt \hrule}}

\def\O#1{O\left(#1\right)}
\def\VEV#1{\left\langle\,#1\,\right\rangle}
\def\tr{\hbox{tr}\,}

\def\be{\begin{equation}}
\def\ee{\end{equation}}
\def\bea{\begin{eqnarray} & &}
\def\eea{\end{eqnarray}}

\def\rf#1{(\ref{#1})}
\def\rfs#1{Eq.~(\ref{#1})}

\def\O {{\cal O}}

\begin{document}
\title{Topological invariants for the fractional quantum Hall states }
\author{Victor Gurarie}
\author{Andrew M. Essin}
\affiliation{Department of Physics, University of Colorado, Boulder,
  CO 80309, USA}


\date{\today}

\begin{abstract}
We calculate a topological invariant, whose value would coincide with the Chern number in case of integer quantum Hall effect, for fractional quantum Hall states.
In case of Abelian fractional quantum Hall states, this invariant is shown to be equal to the trace of the $K$-matrix. In case of  non-Abelian
fractional quantum Hall states, this invariant can be calculated on a case by case basis from the conformal field theory describing these states. This invariant can be used, for example, to distinguish between different
fractional Hall states numerically even though, as a single number, it cannot uniquely label distinct states. 
\end{abstract}

\pacs{67.85.Lm, 03.75.Ss, 67.85.Hj}

\maketitle
It is well known, following the pioneering work of Thouless and collaborators \cite{Thouless1982}, that the Hall conductance of a free fermion system is an integer valued topological invariant, in units of $e^2/h$.  In recent years the concept of topological invariants was generalized to a far larger variety of free electron systems, termed topological insulators \cite{HasanKane2010}, and a plethora of topological invariants for free fermions in a variety of spatial dimensions was proposed and classified \cite{Ryu2010}. These invariants are usually written in a form specifically adapted to noninteracting fermionic systems. For example, they are typically expressed in terms of single particle Bloch waves of the underlying noninteracting Hamiltonians. The topological invariant for the integer quantum Hall effect is the Chern number characterizing the bands of free fermions moving in a two dimensional space. 

At the same time, it is now well understood that these topological invariants can be reexpressed in terms of single particle Green's functions \cite{Niu1985}. In this form, they are defined even if interactions are switched on. Their existence reflects the topology of the Green's functions. However, a topological invariant written in terms of Green's functions no longer corresponds to a response to an external perturbation. For example, the Chern number reexpressed in terms of Green's functions, once the interactions are turned on, is no longer necessarily equal to the Hall conductance (although they remains equal if the interactions are weak \cite{Ludwig2012}, within the integer quantum Hall state). Nevertheless, they retain certain physical meaning thanks to a relationship between the topological invariants in the bulk and at the edge, first derived by G.~Volovik in case of the Chern number type invariant \cite{VolovikBook1}, and subsequently generalized by to a larger class of invariants in a variety of spatial dimensions in Ref.~\cite{EssinGurarie2011}. The edge of a topological insulator is gapless and therefore is not an insulator; the edge invariant is of the type used to characterize topological (semi)-metals such as the one studied in Ref.~\cite{Turner2011}. The precise correspondence between the edge and the bulk invariants is described in Ref.~\cite{EssinGurarie2011}. 

Here we will use this bulk-boundary to calculate the topological invariant, the Chern number reexpressed in terms of single particle Green's functions, in a variety of fractional Hall states, where it is by no means equal to the Hall conductance. We will see that, quite generally, for the states described by a $K$-matrix \cite{Zee1992}, this invariant is equal to the trace of that matrix. 
In a Read-Rezayi non-Abelian state the invariant is equal to $M+2$, where $M$ is defined as in Ref.~\cite{Rezayi1999}. In particular, in the Moore-Read (Pfaffian) state \cite{Read1991} at the filling fraction $\nu=5/2$, $M=1$ and the invariant is equal to 3. Finally, for the anti-Pfaffian state at $\nu=5/2$ the invariant is simply $1$. 

The method we use for calculating the invariant relies on the detailed knowledge one typically has about the low energy theory, and thus the Green's functions, of the edge of fractional Hall states. Indeed, the Green's functions in the bulk are generally not known, thus evaluating the invariant directly in the bulk does not seem to be possible. However, the edge topological invariant relies on the knowledge of the edge Green's functions and can be evaluated directly, which is what we do here. The bulk-boundary correspondence states that the edge and bulk invariant are equal, and thus the edge calculation directly produces the value of the bulk invariant.

The utility of this observation, in our opinion, lies primarily in the possibility of evaluating the invariant numerically. Indeed, if a quantum Hall state candidate is found numerically by exact diagonalization, for example, in order to facilitate its identification one may be able to calculate the Green's functions, and the invariant, in this state. While the invariant does not uniquely label different quantum Hall states as follows from its values listed above, it provides an additional test which helps with the identification of the state found approximately numerically. Note that here we evaluate the invariant by looking at the edge theory of the corresponding fractional Hall state. Numerically it would not be convenient to look at the edge theory directly, rather it is more convenient to construct the theory on a sphere, for example. Then a direct evaluation of the invariant in the bulk becomes possible, which could be compared with the theoretical evaluation  via the edge theory. 

This invariant may also prove useful in the search for fractional quantum Hall analogs in interacting  topological insulators in three spatial dimensions.

To proceed,  let us state the relationship between the bulk invariant and the boundary as it stands in two dimensions. Given the Green's function for an infinite two dimensional system $G_{ab}(\omega,k_x,k_y)$, where $\omega$ is the Matsubara frequency, $k_x$ and $k_y$ are two dimensional momenta, and indices $a$, $b$ refer to the bands and/or spin and flavor of the fermions,  the topological invariant is written as
\be \label{eq:chern} N_2 = \sum_{\alpha \beta \gamma} \epsilon_{\alpha \beta \gamma} \int\! \frac{ d\omega d^2 k}{24 \pi^2}\, \tr G^{-1} \d_{k_\alpha} G G^{-1} \d_{k_\beta} G G^{-1} \d_{k_\gamma} G,
\ee
where $\alpha$, $\beta$ and $\gamma$ are summed over $0$, $1$ and $2$, $k_0 \equiv \omega$, $k_1 \equiv k_x$ and $k_2 \equiv k_y$. The subscript $2$ in $N_2$ indicates that this invariant is for two dimensional systems (suitable generalizations exist in other dimensions). 

This number is an integer regardless of the origin of $G_{ab}(\omega,k_x,k_y)$. In the absence of interactions, $G=\left[ i \omega - {\cal H} \right]^{-1}$, where ${\cal H}$ is the Hamiltonian of noninteracting fermions.  Then $N_2$ is equal to the combined Chern number of the negative energy, single particle bands, and thus coincides with the Hall conductance. In the presence of interactions, it is no longer generally equal to Hall conductance, but retains its topological nature (remains strictly an integer). 

The bulk-boundary correspondence can be stated as
\be \label{eq:bb} N_2 = N_0(\Lambda) - N_0(-\Lambda),
\ee where $N_0$ is the edge topological invariant calculated for the system with a single edge as
\begin{eqnarray} \label{eq:boundary}  N_0(k) =  \int\! \frac{d \omega}{2 \pi i} \int\! dx' dx''\, \tr G^{-1}(\omega, x', x'', k)  \times \cr \d_\omega G(\omega,x'',x',k), &&
\end{eqnarray}
where $G(\omega, x', x'',k)$ is the Green's function of a system with a single edge, which depends on the momentum $k$ along the edge and the two coordinates $x'$ and $x''$ perpendicular to the edge (since the system with an edge is not translationally invariant in the perpendicular direction), and where $\Lambda$ is a suitably chosen (sufficiently large) parameter. For the purpose of \rfs{eq:boundary} the inverse Green's function $G^{-1}$ is defined by
\be \sum_b \int dx' G^{-1}_{ab} (\omega, x, x', k) G_{bc}(\omega, x', x'', k) = \delta_{ac} \, \delta(x-x'').
\ee

The edge invariant $N_0$ encodes the topological information about the edge (in fact, signifies that the edge is a topological metal \cite{VolovikBook1}). For example, in the absence of interactions $N_0(\Lambda) - N_0(-\Lambda)$ is just the number of chiral edge modes, so that  \rfs{eq:bb} simply reflects the fact that the number of chiral edge modes is equal to the bulk conductance $N_2$, something which is well known and can be established in other ways. In the presence of interactions, the meaning of \rfs{eq:bb} is a little less transparent and was established in Ref.~\cite{Gurarie2011} to be the difference between the number of chiral edge modes and chiral edge zeros. 

To further clarify this point,  observe that if one introduces the eigenvalues $\lambda_n$ of $G$, 
\be \sum_b \int dx' G_{ab}(\omega, x, x', k) \psi^{(n)}_b(x') = \lambda_n \psi^{(n)}_a(x),
\ee
then
\be \label{eq:winding} N_0(k) =  \sum_n \int \frac{d\omega}{2\pi i} \d_\omega \ln \lambda_n = \frac{1}{2 \pi } \,  {\rm arg}   \left. \lambda_n  \right|^{\omega = \infty}_{\omega=-\infty}, \ee
which is the sum of windings of the phases of $\lambda_n$ as $\omega$ goes from $-\infty$ to $\infty$ divided by $2\pi$. This last statement can be used to calculate $N_0(k)$ easily. In particular, in the
absence of interactions we have
$\lambda_n = 1/(i \omega - \epsilon_n(k))$, where $\epsilon_n(k)$ are the energy levels of the system with an edge and with momentum $k$ along the edge, and $N_0(k)$
can be calculated as
\be N_0(k) =  \frac{1}{2 \pi } \,  {\rm arg}   \left. \lambda_n  \right|^{\omega = \infty}_{\omega=-\infty} = \oh \sum_n {\rm sign} \, \epsilon_n(k).
\ee
Note that $N_0$ does not evaluate to an integer, because of the slow decay of $\lambda_n$ with $\omega$.  However, differences will still take integer values.
Thus $N_0(\Lambda)-N_0(-\Lambda)$ counts the number of energy levels whose energy 
changes sign as $k$ is varied from $-\Lambda$ to $+\Lambda$. This is simply equal to the number of chiral edge modes.  In the presence of interactions, for example in case of fractional Hall effect, 
the Green's function  eigenvalues are no longer of this simple form. Instead, they generally have not only poles but also zeros as a function of $i \omega$, and $N_0(k)$ counts the difference of the signs of the poles and zeros (see Ref.~\cite{Gurarie2011} for details). 


Let us proceed to calculate the topological invariant \rfs{eq:chern} for a variety of fractional Hall states. As a warm up, let us first consider a simple Laughlin fractional Hall state corresponding to the Hall conductance $\sigma_{xy} = 1/(2m+1)$ and described by a $K$-matrix which reduces to just one number, $K= 2m +1$. The edge Green's function in the position space for such a  state is \cite{Wen1995}
\be \label{eq:grsimp1} G(x,t) \sim \frac{1}{\left( x - v t \right)^{2m+1}},
\ee where $v$ is the velocity of its chiral excitations. Its Fourier transform is given by \cite{Wen1995}
\be \label{eq:wentr} G(\Omega,k) = \int dxdt \,
\frac{e^{i \Omega t- i k x}}{ \left( x - v t \right)^{2m+1} } \sim \frac{\left(\Omega+ v k \right)^{2m}}{ \Omega - v k}.
\ee
This can be obtained, for example, by introducing new variables $s^\pm = x \pm v t$ in the integral. 

We substitute the imaginary frequency $\omega$ into $\Omega = i \omega$ to find
\be  \label{eq:green1} G(\omega, k)  \sim \frac{\left(i \omega+ v k \right)^{2m}}{i  \omega - v k}.
\ee
Then we plug this into \rfs{eq:boundary}.
We can now take advantage of a simple relation, that for any function \be \label{eq:g1} g=\left(i\omega - A\right)^\alpha/\left(i\omega - B \right)^\beta \ee with some real $A$ and $B$,
\be \label{eq:g2} \int_{-\infty}^{\infty} \frac{d \omega}{2 \pi i } g^{-1} \d_\omega g = \oh \left( \beta \, {\rm sign} \, B- \alpha \, {\rm sign} \, A  \right)
\ee [which follows from  \rfs{eq:winding}]. 
Then, putting \rfs{eq:green1} in \rfs{eq:boundary} gives
\be N_0(\Lambda) = \oh \left( 2m + 1\right) \, {\rm sign} \, \Lambda,
\ee
so that
\be N_2 = 2m +1
\ee
from \rfs{eq:bb}.
This is precisely the difference between the number of chiral modes and chiral zeros, represented by the denominator and the numerator of \rfs{eq:green1}. 

First of all, $N_2$ is indeed equal to the trace of $K$-matrix, which in this case is just an odd integer. Also, we see that $N_2$ is by no means equal to 
$\sigma_{xy}$. At the same time, this invariant is identical to the one computed for integer quantum Hall effect with $2m+1$ filled Landau levels. Therefore, 
this invariant is not a unique identifier of a state, and does not necessarily change if a system undergoes a phase transition from an integer to a fractional Hall state. 

Let us now consider more general fractional Hall systems described by a $K$-matrix. Their edge theory is given by the action \cite{Wen1995}
\be \label{eq:actionll}  
S = \sum_{ab} \frac{1}{4\pi} \int dx dt \left[ K_{ab} \d_t \phi_a \d_x \phi_b - V_{ab} \d_x \phi_a \d_x \phi_b \right].
\ee 
Here $V_{ab}$ is the positive definite matrix of velocities of edge excitations, while $K$ is the $K$-matrix, the matrix with integer entries which defines the topological 
order of the fractional Hall state. 
In this representation, we assume the operators creating fermions at the edge are
\be \label{eq:electrons} \psi_a = e^{i \sum_b K_{ab} \phi_b}.
\ee
Let us compute the fermionic Green's function, substitute it into \rfs{eq:bb} and compute $N_2$.

To do that, it is advantageous first to note that since $V$ is a positive definite symmetric matrix, we can parameterize it in terms of some other symmetric matrix $\O$,
\be V = \O^2.
\ee
Changing the variables from $\phi$ to $\varphi = \O \phi$ (which now has units of $\sqrt{\text{velocity}}$), we find the new action
\be S = \frac{1}{4\pi}  \int dx dt \left[ \d_t \varphi \, \O^{-1} K \O^{-1}\d_x \varphi - \d_x \varphi \d_x \varphi \right].
\ee 
Here the matrix notation for matrix products is used, for brevity. 
It is now convenient to diagonalize the symmetric matrix 
\be \label{eq:rel} \O^{-1} K \O^{-1} = U^T \Pi U,\ee 
where $\Pi$ is a diagonal matrix, and $U$ is an orthogonal matrix, $U^T U=1$. 
Another change of variables
\be \tilde \varphi =  U \varphi
\ee 
brings the action to the simple form
\be S = \frac{1}{4\pi} \sum_a \int dx dt \left[   \d_x \tilde \varphi_a \d_t \tilde \varphi_a/v_a - \left( \d_x \tilde \varphi_a \right)^2 \right] .
\ee 
Here $v_a$ are the inverses of the diagonal entries of the diagonal matrix $\Pi$, which obviously have units of velocity.
It is straightforward to calculate the correlation functions of $\tilde \varphi$ now. They are given by
\be \VEV{ \tilde \varphi_a(x,t) \, \tilde \varphi_b(0,0)} = - 2\pi i \, \delta_{ab} \int \frac{d\omega dk}{(2\pi)^2} \frac{e^{i kx - i \omega t}}{ \omega k/v_a - k^2}.
\ee
Doing the integral results in the expression
\be \VEV{\tilde \varphi_a(x,t) \, \tilde \varphi_b(0,0)} = \left| v_a \right| \ln \frac{L}{x-v_a t},
\ee where $L$ is the system size. 

Now we are interested in calculating
\begin{eqnarray} G_{ab} &=& \VEV{\psi_a(x,t) \, \psi^\dagger_b(0,0)}= \cr && \VEV{e^{i \sum_c K_{ac} \phi_c(x,t)} \, e^{-i \sum_d K_{bd} \phi_d(0,0)}},
\end{eqnarray} as follows from \rfs{eq:electrons}. To compute that, we 
work out $\phi$ in terms of $\tilde \varphi$ to find
\begin{eqnarray} &&  \VEV{ \left( \sum_c K_{ac} \phi_c(x,t) \right) \, \left( \sum_d K_{bd} \phi_d(0,0) \right)} =\cr && \left[ \O U^T \left| \Pi \right|  \ln \left( \frac{L}{x-t/\Pi} \right) U \O\right]_{ab}
\end{eqnarray}
where the expression in the square brackets is understood as a product of matrices, and $\left| \Pi \right| \ln \left(L/(x-t/\Pi) \right)$ is understood as  a diagonal matrix with the
diagonal 
entries $ \ln \left( L/(x- v_a t) \right)/\left| v_a \right|$. 
Finally, using the standard formula of Gaussian integration
\be \label{eq:ga} \VEV{e^{i \phi_a(x,t)} \, e^{-i \phi_b(0,0)}} = e^{- \oh \left[\phi_a(x,t) - \phi_b(0,0) \right]^2}
\ee
we find
\be \label{eq:fgr} G_{ab}(x,t) \sim \prod_c \frac{\delta_{ab}}{\left( x-v_c t \right)^{\frac{1}{\left| v_c \right|} \left(\sum_d U_{cd} \O_{da} \right)^2}}.
\ee
A word must be said about why the off-diagonal terms in $G_{ab}$ are zero. Evaluating these terms via \rfs{eq:ga} gives, for their $L$-dependence,
\be G_{ab} \sim L^{-\oh \sum_c \left[ \sum_d  \left( U_{cd} \O_{da} - U_{cd} \O_{db} \right) \right]^2 \frac{1}{ \left| v_c \right|}}.
\ee
If $a \not = b$ the exponent is negative (columns of $U\mathcal{O}$ are linearly independent), so this goes to zero as $L$ goes to infinity, justifying the Kronecker delta in \rfs{eq:fgr}. If $a=b$, this is equal to 1 and $G_{aa}$ does not depend on $L$ as indicated in \rfs{eq:fgr}. \rfs{eq:fgr} generalizes
the Green's functions of fermions  given by \rfs{eq:grsimp1} to  the case of a generic $K$-matrix. 

We now need to perform a Fourier transform $G_{ab}(x,t)$. This is hard to do in general, and we expect that the answer is not as simple as it was in the 
previously considered case \rfs{eq:wentr}. However, all we are interested in is a change in the argument of $G_{aa}(\omega,k)$  as $\omega$ is varied from $-\infty$ to
$\infty$, as explained 
in \rfs{eq:winding}. This can be found in a relatively straightforward way.

Let us calculate the Fourier transform of \rfs{eq:fgr} by introducing the imaginary time $\tau = i t$ directly, 
\be \label{eq:grsimp} G_{aa}(\omega,k)=  \int  d\tau dx  \, \frac{e^{i \omega \tau - i k x}}{\prod_c \left( x + i v  x_c \tau \right)^{\alpha_c^a}}.
\ee 
Here  $\alpha_c^a = \frac{1}{\left| v_c \right|} \left(\sum_d U_{cd} \O_{da} \right)^2>0$, $v>0$ is some positive constant with the units of velocity, and $x_a=v_a/v$ are some numbers of an arbitrary sign. 
We are then interested in calculating the change in the argument of $G$ as $k$ is kept fixed and $\omega$ is swept from $-\infty$ to $\infty$. Let us show that
this change is equal to 
\be \label{eq:change}
N^a_0(k)= \frac{1}{2\pi } \left. \arg G_{aa}(\omega,k)  \right|_{\omega=-\infty}^{\omega=\infty} =\frac{m_a}{2} \, {\rm sign} \, k,
\ee 
where
$N^a_0(k)$ is the contribution to $N_0(k)$ from $G_{aa}$ [it needs to be summed over $a$ to find $N_0(k)$] and $m_a$ is defined as
\be \label{eq:m} m_a= \sum_c \alpha_c^a \, {\rm sign} \, x_c = \left[ \O U^T \Pi U \O \right]_{aa}=K_{aa}.
\ee
To show that this is a diagonal entry of $K$, we took advantage of \rfs{eq:rel}. $m_a$ is an integer since $K$ is integer valued.

Now we go back to evaluating \rfs{eq:grsimp}. We change the variables from $\tau$, $x$ to 
\be x+i v \tau = r e^{i \phi}.
\ee
This gives (for brevity, we suppress the index $a$ in the Green's function and as an upper index of $\alpha$)
\be G = \int_0^\infty \frac{r dr}{v r^{n}} \int_{-\pi}^\pi  d\phi \frac{e^{i  q r \cos(\phi-\phi_0)}}{\prod_c \left( \cos(\phi) + i x_c \sin(\phi) \right)^{\alpha_c}},
\ee
Here $\phi_0$ is the angle between the 
vectors $(-k, \omega/v)$ and the vector $(1,0)$, $r= \sqrt{x^2 + v^2 \tau^2}$, $q = \sqrt{ k^2 + \omega^2/v^2}$, and $n=\sum_a \alpha_a$. We rewrite this integral
as
\be G = \int\! \frac{ r dr}{v r^{n}} \int_{-\pi}^\pi\! d\phi \frac{e^{i  q r \cos \phi}}{\prod_c \left( e^{i (\phi+\phi_0)} \frac{1+x_c}{2} + e^{- i ( \phi+\phi_0)} \frac{1-x_c}{2} \right)^{\alpha_c}}.
\ee 
Further analysis depends on whether $x_c$ are positive or negative. If $x_c$ is positive, we take $e^{i(\phi+\phi_0)} (1+x_c)/2$ outside the appropriate bracket, otherwise
we do the same with $e^{-i (\phi+\phi_0)} (1-x_c)/2$. 
We find
\be \label{eq:greenexplicit} G \sim e^{-i m \phi_0} \int \frac{ r dr}{v r^{n}} \int_{-\pi}^\pi d\phi \frac{e^{i q r \cos(\phi)  - i m \phi}}{\prod_c \left( 1+ e^{-2 i (\phi+\phi_0) \, {\rm sign} \, x_c} \rho_c \right)},
\ee
where $m$ was defined above in \rfs{eq:m} and $\rho_c = (1-|x_c|)/(1+|x_c|)$, $| \rho_c | <1$. 

As $\omega$ is swept from $-\infty$ to $\infty$, $\phi_0$ grows from $-\pi/2$ to $\pi/2$ if $k<0$ and decreases from $3\pi/2$ to $\pi/2$ if $k>0$. We would like to know the change in the argument of $G$ during this process. The outer explicit factor of $e^{-i m \phi_0}$ obviously contributes $\pi m \, {\rm sign} \, k$ to that change. Let 
us argue that the rest of the expression does not contribute at all to the change of the argument. To do that, we rewrite everything which multiplies $e^{-i m \phi_0}$ in this expression as
\be \label{eq:zzz}  \int \frac{r dr}{v r^{n}} \int_{-\pi}^\pi d\phi \frac{e^{i q r \cos(\phi)  - i m \phi}}{\prod_c \left( 1+ e^{-2 i (\phi+\phi_0) \, {\rm sign} \, x_c} \rho_c \right)^{\alpha_c}}.
\ee
It is clear that if all $\rho_c=0$ (which corresponds to all velocities being equal to each other in magnitude and equal to $v$), then this expression is independent of $\phi_0$ and its argument does not wind at all as a function of $\phi_0$. As $\rho_c$ are increased, in order for the argument to start winding, one should be able to find such $\phi_0$, at some $\rho_c$, that this expression is equal to zero (or infinity, but this expression is finite as long as all $\left| \rho_c \right| <1$. 

However, if this expression is zero, that means that the Green's function has a zero at those values of $\rho_c$. Zeros of the Green's function can occur only at $\omega=0$ (see page 168 of Ref.~\cite{AGDBook} or Ref.~\cite{Gurarie2011} for a slightly stronger version of this statement with a more detailed discussion). $\omega=0$ corresponds to $\phi_0=0$ or $\phi_0=\pi$. However, if such a zero were to occur, then for a related system with velocities $v_a' = v/v_a$ where $\rho_c' = -\rho_c$, the zeros would occur at $\phi=\pi/2$ which contradicts the theorem that zeros occur only at $\omega=0$ or $\omega=\pi$. Therefore, the end result is that this is impossible. We conclude from this that \rfs{eq:zzz} does not wind around zero of the complex plane as $\phi_0$ changes from $-\pi/2$ to $\pi/2$, or from $3\pi/2$ to $\pi/2$. 

Taken together, this shows that indeed the contribution to $N_0(k)$ from $G_{aa}$ is given by Eqs.~\rf{eq:change} and \rf{eq:m} to be $K_{aa}$. Summing over all the entries of the Green's function we find that
\be N_0(k) = \oh \, {\rm sign} \, k \, \cdot  \tr K .
\ee
That, in turn, leads to the anticipated answer,
\be N_2 = N_0(\Lambda) - N_0(-\Lambda) = \tr K.
\ee

Note that this expression is not invariant under the change of basis (going from the basic fields $\phi_a$ in Eq.~\rf{eq:actionll} to some linear combination thereof). Indeed, the topological invariant
is defined in the preferred basis where the fermion operators are simply defined, as in \rfs{eq:electrons}. It is possible to rewrite the expression for the topological invariant in the basis independent form with the help of the ``electron lattice'' matrix $C$ introduced in Ref.~\cite{Moore2002}; since, in the present context, the purpose of this matrix would just be to take the matrix $K$ back to the preferred ``electron" basis, we omit this.


Now, let us now briefly look at the non-Abelian Read-Rezayi states \cite{Rezayi1999} (which include the Pfaffian, or Moore-Read, state as a particular case \cite{Read1991}). 
The edge theory of these states include one charged and one neutral particle (a Luttinger liquid-type boson and a parafermion, which reduces to a Majorana fermion in case of the Moore-Read state). The Green's functions at the boundary can be easily derived from conformal field theory and are equal to
\be \label{eq:rr} G \sim \frac{1}{\left( x - v_n t \right)^{2- \frac 2 N} \left( x- v_c t \right)^{M+ \frac 2 N} }.
\ee
Here $N$ is the level of the Read-Rezayi state ($N=2$ corresponds to the Pfaffian state) and $M$ is an odd positive integer (even $M$ would correspond to a bosonic state which we do not discuss here). The filling fraction of these states  is known to be \cite{Rezayi1999}
\be \nu = \frac{N}{M N + 2}
\ee
(this number represents the filling fraction of the partially filled Landau level; in the presence of lower completely filled Landau levels the filling fraction can be larger than this number by an integer).

It is now straightforward to use  Eqs.~\rf{eq:grsimp} and \rf{eq:change} to find that the topological invariant is the sum of two exponents in \rfs{eq:rr}, or
\be N_2 = M+2.
\ee 
Interestingly, it is independent of $N$. For the $5/2$ plateau in fractional Hall effect, understood as $M=1$, $N=2$ Pfaffian state, this gives $N_2=3$, the same as for the Laughlin $\nu=1/3$ state. 

A final non-Abelian  state of interest to us is the anti-Paffian state at $\nu=5/2$ \cite{Levin2007,Lee2007,Bishara2008}. Its edge Green's function is given by
\be  \label{eq:apf} G \sim \frac{1}{\left( x- v_c t \right)^{2} \left(x+v_n t \right)}.
\ee Importantly, here the charged and neutral modes are counterpropagating. As a result, employing  Eqs.~\rf{eq:grsimp} and \rf{eq:change} we find that the topological invariant is the {\sl difference} of the two exponents in \rfs{eq:apf} or
\be N_2 = 1,
\ee the same as in the simple $\sigma_{xy}=e^2/h$ integer quantum Hall state. Note that in all these cases, the invariant $N_2$ appears simply to reflect the scaling dimension of the electron operator \cite{Wen1995}. 

While the value of the invariant would not be helpful in trying to distinguish the anti-Pfaffian from a simple integer Hall state, it can serve to distinguish the Pfaffian ($N_2=3$) from the anti-Pfaffian ($N_2=1$) $\nu=5/2$ state. 

In conclusion, we have defined an invariant $N_2$ for fractional Hall states via \rfs{eq:chern} and calculated it for a variety of fractional and non-Abelian quantum Hall states. 
We find that the invariant is not a unique identifier of the state; a phase transition can occur while the invariant may stay the same. However, within the standard low-energy descriptions employed here, it appears that the invariant cannot change without a phase transition since its value is defined by the state; this should be contrasted with the situation in one dimensional systems where the invariant can change without a phase transition \cite{Gurarie2012}. 

We would like to add one final remark: it is possible to try to define the invariant not via the Green's functions but rather via the phases (boundary conditions) across the system \cite{Niu1985} which can lead to an expression equal to the 
fractional Hall conductance, unlike the invariant constructed here, which is always an integer and is not equal to the Hall conductance.  This may be an interesting avenue to pursue in the future.

VG would like to acknowledge NSF grant no.~DMR-1205303 for support, and is grateful to A. Altland for useful discussions, in particular for pointing out the diagonal structure of the Green's functions in \rfs{eq:fgr}. 

\bibliography{FQHE.bib}
\end{document}